%% file: main.tex
\pdfoutput=1

\documentclass[11pt]{article}

\usepackage[final]{acl}  

\input{preamble/packages}

\input{preamble/definitions}

\title{Optimized Text Embedding Models and Benchmarks\\ for Amharic Passage Retrieval}

\input{preamble/authors}

\begin{document}

\maketitle
\begin{abstract}
Neural retrieval methods using transformer-based pre-trained language models have advanced multilingual and cross-lingual retrieval. However, their effectiveness for low-resource, morphologically rich languages such as Amharic remains underexplored due to data scarcity and suboptimal tokenization. We address this gap by introducing Amharic-specific dense retrieval models based on pre-trained Amharic BERT and RoBERTa backbones. Our proposed \textit{RoBERTa-Base-Amharic-Embed} model (110M parameters) achieves a 17.6\% relative improvement in MRR@10 and a 9.86\% gain in Recall@10 over the strongest multilingual baseline, \textit{Arctic Embed 2.0} (568M parameters). More compact variants, such as \textit{RoBERTa-Medium-Amharic-Embed} (42M), remain competitive while being over 13$\times$ smaller. Additionally, we train a \textit{ColBERT-based late interaction retrieval model} that achieves the highest MRR@10 score (0.843) among all evaluated models. We benchmark our proposed models against both sparse and dense retrieval baselines to systematically assess retrieval effectiveness in Amharic. Our analysis highlights key challenges in low-resource settings and underscores the importance of language-specific adaptation. To foster future research in low-resource IR, we publicly release our dataset, codebase, and trained models.\footnote{\url{https://github.com/kidist-amde/amharic-ir-benchmarks}}
\end{abstract}

\input{sections/introduction}

\input{sections/Motivation}
\input{sections/Related-work}
\input{sections/Benchmark_model}

\input{sections/Experimental}

\input{sections/Results}
\input{sections/Conclusion}
\input{sections/acknowledgements}

\clearpage
\input{sections/Limitations}

\input{sections/Ethical_considerations}

\bibliography{references}

\input{sections/Appendix}

\end{document}

%% file: preamble/packages.tex
\usepackage{times}
\usepackage{latexsym}
\usepackage{multirow}
\usepackage{booktabs}

\usepackage[T1]{fontenc}

\usepackage[utf8]{inputenc}

\usepackage{microtype}

\usepackage{inconsolata}

\usepackage{graphicx}

\usepackage[inline]{enumitem}
\usepackage{xcolor,colortbl}
\usepackage{acronym}
\usepackage{bm}
\usepackage[skip=1mm]{caption}
\usepackage{natbib}
\usepackage{hyperref}
\usepackage{amsmath}
\usepackage{xcolor}  

%% file: preamble/definitions.tex
\definecolor{Gray}{gray}{0.85}
\definecolor{LightCyan}{rgb}{0.88,1,1}

\makeatletter
\def\heading{\@startsection{paragraph}{4}{\z@}{.1ex plus 0.1ex minus 0.1ex}{-.25em}{\normalsize\bf}}
\makeatother

\AtBeginDocument{%
  }

\acrodef{DDRO}{direct document relevance optimization}
\acrodef{IR}{information retrieval }
\acrodef{DSI}{differentiable search indexes}
\acrodef{NLP}{natural language processing}
\acrodef{GenIR}{generative information retrieval}
\acrodef{LM}{language model}
\acrodef{RLRF}{reinforcement learning from relevance feedback}
\acrodef{SFT}{supervised fine-tuning}
\acrodef{ANN}{approximate nearest neighbor}
\acrodef{LSR}{learned sparse retrieval}

\hypersetup{
  colorlinks   = true,    
  urlcolor     = blue,    
  linkcolor    = blue,    
  citecolor    = blue      
}

%% file: preamble/authors.tex
\author{
  Kidist Amde Mekonnen\thanks{Equal contribution.} \\
  University of Amsterdam \\
  \texttt{k.a.mekonnen@uva.nl}
  \And
  Yosef Worku Alemneh\footnotemark[1] \\
  Independent Researcher \\
  \texttt{yosefwalemneh@gmail.com}
  \And
  Maarten de Rijke \\
  University of Amsterdam \\
  \texttt{m.derijke@uva.nl}
}

%% file: sections/introduction.tex
\section{Introduction}
\label{intro}
As a foundational task in \ac{NLP}, document retrieval plays a crucial role in applications such as open-domain question answering~\cite{chen2017reading} and fact-checking~\cite{thorne2018fever}.
Traditional retrieval systems such as TF-IDF and BM25~\cite{Robertson1997OnRW, Robertson2009ThePR} match queries to documents based on lexical overlap. While efficient, they struggle with vocabulary mismatch and semantic ambiguity, limiting their generalizability to synonyms and paraphrases. These challenges are particularly pronounced in morphologically rich languages, where high inflectional variability and complex morphology complicate exact-match retrieval. Suboptimal tokenization in multilingual models further exacerbates these issues, leading to over-segmentation and inefficient subword representations~\cite{rust-etal-2021-good}. As a result, word-based indexing methods fail to capture non-concatenative morphology, affixation, and orthographic variations, degrading retrieval effectiveness. To address these limitations, retrieval models must move beyond lexical overlap and incorporate robust semantic representations.

\heading{Neural retrieval models.}
Recent work has introduced several families of neural retrieval methods that leverage transformer-based pre-trained language models to improve retrieval effectiveness, particularly in monolingual English settings. These methods have significantly advanced document ranking, achieving state-of-the-art performance in benchmarks such as MS MARCO~\cite{Campos2016MSMA} and Natural Questions~\cite{kwiatkowski-etal-2019-natural}. Broadly, they fall into three main categories~\cite{yates-etal-2021-pretrained}: 
\begin{enumerate*}[label=(\roman*)]
\item learned sparse retrieval~\cite[e.g., SPLADE,][]{formal2021splade-v2}, which enhances queries and documents with \mbox{context}-aware term expansions; 
\item dense retrieval~\cite[e.g., DPR,][]{karpukhin-etal-2020-dense}, which maps text into dense vector spaces for efficient retrieval, employing a dual-encoder architecture that encodes queries and documents separately, a design that limits their effectiveness for fine-grained relevance modeling; 
and
\item cross-encoders~\cite[e.g.,][]{nogueira2019passage, nogueira2019multi}, which address this limitation by jointly encoding query-document pairs, capturing richer contextual interactions, with a computational overhead that restricts their use to re-ranking candidate documents~\cite{humeau2019poly}. 
As an alternative, late-interaction models~\cite[e.g., ColBERT,][]{Khattab2020ColBERTEA}, introduce token-level interactions and strike a balance between the efficiency of dense retrieval and the expressiveness of cross-encoders.
\end{enumerate*}

A newer paradigm, \acl{GenIR}~\citep{ Metzler2021RethinkingS,Tay2022TransformerMA,chen-2023-unified}, uses pre-trained encoder-decoder models to consolidate indexing, retrieval, and ranking into a single generative framework. While promising, GenIR lags behind dense retrieval in handling large-scale datasets and accommodating dynamic corpora, requiring further study of its scalability and adaptability~\cite{pradeep-etal-2023-generative}.

\heading{Research gap.}
Despite these advances, neural retrieval remains understudied for morphologically complex, low-resource languages like Amharic. Most retrieval models are optimized for high-resource languages, and prior work has largely focused on cross-lingual transfer from these languages~\cite{GreenLLM}. Despite advancements in multilingual embedding models~\cite{wang2024multilingual, yu2024arctic}, these approaches remain inadequate for morphologically rich languages due to suboptimal tokenization, poor subword segmentation, and weak cross-lingual transfer~\cite{ustun-etal-2019-cross}. Section~\ref{motive} further explores the importance of addressing this gap in information retrieval research.

\heading{Our contribution.}
To address the gap identified above, we focus on Amharic and introduce optimized retrieval models and benchmarks, making the following key contributions:
\begin{enumerate*}[label=(\roman*)] 
\item \emph{Amharic text embeddings}: we develop dense retrieval models for Amharic, leveraging Amharic BERT and RoBERTa as base models, improving passage ranking accuracy for morphologically complex text.   

\item \emph{The first systematic benchmark for Amharic}: we evaluate both sparse and dense retrieval models on Amharic, establishing strong baselines for future research.

\item \emph{A language-specific vs.\ multilingual analysis}: we show that Amharic-optimized models consistently outperform multilingual embeddings, underscoring the value of language-specific adaptation.  

\item \emph{A public benchmark dataset}: We repurpose the Amharic News Text Classification Dataset (AMNEWS) by treating headlines as queries and corresponding articles as passages, creating MS MARCO-style query-passage pairs with heuristic relevance labels. This enables reproducible evaluation of passage ranking models for Amharic. We refer to this processed version as the \emph{Amharic Passage Retrieval Dataset}. The dataset is publicly available on Hugging Face,\footnote{\url{https://huggingface.co/datasets/rasyosef/amharic-news-retrieval-dataset}} and all code and preprocessing scripts are released on GitHub.\footnote{\url{https://github.com/kidist-amde/amharic-ir-benchmarks/tree/main/data}}

\end{enumerate*}

%% file: sections/Motivation.tex
\section{Motivation}
\label{motive}
Recent studies highlight systemic shortcomings in low-resource language technologies, leading to retrieval failures, biased outputs, and exposure to harmful or policy-violating content~\cite{shen-etal-2024-language, nigatu2024searched}. For example, \citet{nigatu2024searched} find that Amharic-speaking YouTube users frequently encounter such content due to retrieval systems misinterpreting user intent behind benign queries. These errors stem from foundational limitations in \ac{IR} systems, which are optimized for high-resource languages like English and struggle with morphologically complex languages like Amharic. 
The consequences extend beyond search engines: \citet{sewunetie-etal-2024-gender} demonstrate that retrieval failures in machine translation propagate gender bias, defaulting Amharic occupational terms to male forms even when the context is gender-neutral. Such errors reflect broader research gaps in \ac{NLP}, where systems disproportionately prioritize high-resource languages, thereby exacerbating inequities faced by underrepresented linguistic communities~\cite{shen-etal-2024-language}.

Amharic, the working language of Ethiopia’s federal government and one of the most widely spoken Semitic languages~\cite{gezmu-etal-2018-contemporary}, presents unique challenges for \ac{IR}. Its root-based templatic morphology allows a single root to generate numerous derived forms through affixation and vowel pattern changes. These morphological variations, combined with the Ge’ez script, an Abugida writing system with \texttt{33} base characters and over \texttt{230} syllabic forms, make Amharic structurally and morphologically distinct from Indo-European and other high-resource languages. As a result, conventional retrieval models tend to underperform without language-specific adaptation. Addressing these challenges requires Amharic-specific embedding models tailored for passage retrieval. While recent efforts~\cite{destaw-etal-2021-development, azime-etal-2024-walia} have advanced Amharic \ac{NLP}, their primary focus has not been on optimizing retrieval performance.

Our work fills this gap by developing and benchmarking retrieval methods specifically adapted to Amharic’s linguistic characteristics, laying a foundation for more equitable and semantically accurate information access in low-resource language settings.

%% file: sections/Related-work.tex
\section{Related Work}
\label{related}  
Retrieval systems commonly adopt a two-stage pipeline to optimize efficiency and effectiveness:   
\begin{enumerate*}[label=(\roman*)]  
\item First-stage retrieval efficiently retrieves candidate documents using lightweight methods such as sparse or dense retrieval.  
 
\item Re-ranking refines the results using computationally more intensive models, such as cross-encoders.  
\end{enumerate*}  

\heading{Sparse retrieval.}  
Sparse retrieval is fundamental in IR, with BM25 known for its efficiency, interpretability, and cross-domain robustness~\cite{Robertson2009ThePR}. However, it struggles with vocabulary mismatch and morphological variability, challenges that are particularly acute in morphologically rich languages like Amharic. \Ac{LSR} methods~\cite{formal2021splade, formal2021splade-v2} attempt to mitigate these issues by dynamically weighting and expanding terms, thereby enhancing relevance while maintaining interpretability~\cite{dai2020context}. However, LSR faces limitations in low-resource settings due to the scarcity of annotated data, dialectal diversity, and morphological complexity (e.g., Amharic's templatic morphology), which necessitate subword-aware tokenization or morphological analyzers that are often unavailable.

\heading{Dense retrieval.} Dense retrieval encodes queries and documents into a shared semantic space using neural network encoders, enabling efficient retrieval via \ac{ANN} search based on embedding similarity~\cite{johnson2019billion, karpukhin-etal-2020-dense, Xiong2020ApproximateNN}. While it helps mitigate lexical mismatch, its effectiveness in low-resource languages is hindered by the need for large-scale labeled training data. Multilingual models such as mBERT~\cite{mbert}, XLM-R~\cite{XLM-R}, and African language-specific models like SERENGETI~\cite{SERENGETI} and AfriBERTa~\cite{AfriBERTa} partially address data scarcity through cross-lingual pretraining. However, their effectiveness in morphologically complex languages like Amharic has not been thoroughly investigated.

Recent advances in unsupervised contrastive learning, such as Contriever~\cite{Izacard2021UnsupervisedDI}, have demonstrated strong zero-shot and multilingual retrieval performance, especially in cross-lingual transfer scenarios. Nonetheless, their effectiveness in morphologically complex languages like Amharic remains unexplored, as current evaluations do not account for challenges arising from root-based and templatic morphologies.

Beyond data scarcity, retrieval performance is further constrained by morphological complexity and tokenization challenges. Amharic’s templatic morphology often causes standard subword tokenizers to over-segment words into non-morphemic units, leading to fragmented representations that obscure semantic relationships. Broader research on multilingual tokenization quality~\cite{rust-etal-2021-good} shows that excessive segmentation in morphologically rich languages introduces noise into subword representations, degrading performance in downstream tasks.

Despite recent advances in multilingual dense retrieval, state-of-the-art models such as Arctic Embed 2.0~\cite{yu2024arctic} and Multilingual E5~\cite{wang2024multilingual}, which topped the \textit{MTEB Embedding Leaderboard}\footnote{\url{https://huggingface.co/spaces/mteb/leaderboard}} at the time of our study, continue to struggle with highly inflected languages. These models often produce suboptimal tokenizations, fragmented subword representations, and inefficient embeddings, ultimately limiting their retrieval effectiveness. Our empirical findings in Section~\ref{sec:Fertility} illustrate the extent to which tokenization errors impair retrieval performance in Amharic.


\heading{Bridging the gap in Amharic IR.}   
Retrieval systems are primarily optimized for high-resource languages, exacerbating performance disparities in low-resource settings like Amharic~\cite{nigatu2024searched}. Prior research in Amharic \ac{IR} has explored pre-trained embeddings~\cite[Word2Vec, fastText, AmRoBERTa,][]{destaw-etal-2021-development}, morphological tools~\cite[e.g., annotation frameworks, WordNet-based query expansion,][]{yeshambel2021morphologically}, and cross-lingual transfer via multilingual models~\cite[AfriBERTa,][]{azime2024enhancing}. However, systematic evaluations of sparse and dense retrieval architectures remain absent, making principled comparisons difficult and leaving the effectiveness of different paradigms in Amharic IR largely unexamined.

\citet{2AIRTC} introduce 2AIRTC, a TREC-style test collection for standardized Amharic IR evaluation, but it lacks baseline retrieval benchmarks and complete relevance judgments, making recall-based assessments unreliable. To ensure robust evaluation, we conduct our main experiments on the Amharic Passage Retrieval Dataset, which we derive by preprocessing the Amharic News Text Classification Dataset (AMNEWS)~\cite{am_news_data} into MS MARCO-style query-passage pairs (see Section~\ref{exp}). A detailed analysis of 2AIRTC, its limitations, and our supplementary evaluations on this dataset is provided in Appendix~\ref{sec:appendix}.

To address these gaps, our work introduces Amharic-specific retrieval models that incorporate both strong and compact encoder backbones (Section~\ref{sec:amharic_embedding_models}), optimized using contrastive training to better handle Amharic’s morphological complexity. We also develop and evaluate a late-interaction ColBERT model tailored for Amharic, and benchmark both sparse and dense retrieval architectures. This enables rigorous, reproducible comparisons across retrieval paradigms.

%% file: sections/Benchmark_model.tex
\section{Methodology}
\label{sec:methodology}
In this section, we outline our approach to Amharic dense retrieval. We begin by reviewing dense retrieval and ColBERT architectures, which underpin our framework. We then introduce our Amharic embedding models, describing their architecture, training setup, and optimization strategy.

\subsection{Preliminaries}
\label{sec:preliminaries}

\subsubsection*{Dense retrieval models}
Dense retrieval maps queries and passages into a shared vector space using transformer-based encoders~\cite{karpukhin-etal-2020-dense}. Given a query \( q \) and a set of candidate passages \( P = \{ p_1, p_2, ..., p_N \} \), a dense retrieval model maps each input to a fixed-length vector representation via a transformer encoder \( \text{Enc}(\cdot) \):
\begin{equation}
    q_{\text{enc}} = \text{Enc}_Q(q), \quad p_{\text{enc}} = \text{Enc}_P(p)
\end{equation}
The relevance score between a query \( q \) and a passage \( p \) is computed using a similarity function \( f(q, p) = \text{sim}(q_{\text{enc}}, p_{\text{enc}}) \), where \( \text{sim}(\cdot, \cdot) \) typically denotes the dot product or cosine similarity.

\subsubsection*{ColBERT: Late interaction retrieval}
ColBERT~\cite{Khattab2020ColBERTEA} enhances retrieval by preserving token-level interactions between queries and passages. Rather than aggregating inputs into a single vector, it encodes:
\begin{equation}
\mbox{}\hspace*{-2mm}
\resizebox{0.92\linewidth}{!}{$
    q_{\text{enc}} = [\mathbf{h}_q^1, \mathbf{h}_q^2, \ldots, \mathbf{h}_q^m], \ p_{\text{enc}} = [\mathbf{h}_p^1, \mathbf{h}_p^2, \ldots, \mathbf{h}_p^n]
    $}
\hspace*{-2mm}\mbox{}
\end{equation}
where \( \mathbf{h}_q^i \) and \( \mathbf{h}_p^j \) are contextualized token embeddings. Relevance is computed using maximum similarity pooling:
\begin{equation}
    f(q, p) = \sum_{i=1}^{m} \max_{j \in \{1, \dots, n\}} \text{sim}(h_q^i, h_p^j).
\end{equation}
This allows fine-grained token-level matching while remaining efficient at inference time.

\subsection{Amharic Text Embedding Models}
\label{sec:amharic_embedding_models}
We design three transformer-based dense retrieval models for Amharic, each with different parameter sizes. All models use a context length of \texttt{512} tokens to balance effectiveness and efficiency.

\begin{enumerate}[label=(\arabic*), leftmargin=*, nosep]
    \item \textbf{RoBERTa-Base-AM-Embed} (\texttt{110M} parameters): A 12-layer transformer with hidden size \texttt{768}, based on XLM-RoBERTa~\cite{XLM-R}. This model offers deep contextualized representations while remaining compatible with standard retrieval pipelines.
    
    \item \textbf{RoBERTa-Medium-AM-Embed} (\texttt{42M} parameters): A compact 8-layer transformer with hidden size \texttt{512}, optimized for retrieval latency and resource-constrained environments.
    
    \item \textbf{BERT-Medium-AM-Embed} (\texttt{40M} parameters): Based on the original BERT architecture~\cite{devlin2018bert}, with 8 layers and hidden size \texttt{512}. This model is suited for latency-sensitive applications.
\end{enumerate}

\heading{Embedding Vector Generation:} To obtain passage embeddings, we apply the following steps to the last hidden states of the pre-trained Amharic base models:
\begin{enumerate}[label=(\roman*)]
    \item \textbf{Mean pooling:} Aggregate token embeddings to form a fixed-length vector:  
    \[
    \mathbf{h}_{\text{pool}} = \frac{1}{T} \sum_{t=1}^{T} \mathbf{h}_t
    \]
    where \( T \) is the sequence length.
    
    \item \textbf{L2 normalization:} Normalize embeddings to unit length for cosine similarity:
    \[
    \mathbf{h}_{\text{norm}} = \frac{\mathbf{h}_{\text{pool}}}{\|\mathbf{h}_{\text{pool}}\|_2}
    \]
\end{enumerate}

\heading{Training setup.} All models are initialized from Amharic pre-trained checkpoints (Amharic BERT and RoBERTa) and fine-tuned using contrastive learning with in-batch negatives on a corpus of \texttt{45K} Amharic query-passage pairs. Models are trained for 4 epochs using the AdamW optimizer (\texttt{lr} = 5e-5) with cosine learning rate decay. We evaluate using MRR, NDCG, and Recall@K. Passages longer than \texttt{512} tokens are truncated. Additional implementation details are in Section~\ref{details}.

\heading{Multiple negatives ranking loss (MNRL).} Following~\cite{reimers-2019-sentence-bert}, we use in-batch negatives to train our models. For a batch of queries \(\{\mathbf{q}_i\}_{i=1}^B\), their corresponding positives \(\{\mathbf{p}_i^+\}_{i=1}^B\), and in-batch negatives \(\mathcal{N}_i = \{\mathbf{p}_j\}_{j \neq i}\), the loss \(\mathcal{L}\) is:
\begin{equation}
\mbox{}\hspace*{-2mm}
\resizebox{0.92\linewidth}{!}{$
\mathcal{L} = -\frac{1}{B} \sum_{i=1}^B \log \frac{\exp(f(\mathbf{q}_i, \mathbf{p}_i^+))}{\exp(f(\mathbf{q}_i, \mathbf{p}_i^+)) + \sum_{\mathbf{p}_j^- \in \mathcal{N}_i} \exp(f(\mathbf{q}_i, \mathbf{p}_j^-))}
$}
\hspace*{-2mm}\mbox{}
\end{equation}
This loss encourages the model to assign higher similarity scores to the relevant passages \(\mathbf{p}_i^+\) relative to the in-batch negatives \(\mathcal{N}_i\), promoting discriminative representations in the shared embedding space.

%% file: sections/Experimental.tex
\section{Experimental Setup}
\label{exp}

\subsection{Training Data}
We conduct our experiments using the Amharic Passage Retrieval Dataset, which we construct by preprocessing the Amharic News Text Classification Dataset (AMNEWS)~\cite{am_news_data}. The original dataset contains \texttt{50,706} Amharic news articles categorized into six domains: Local News, Sports, Politics, International News, Business, and Entertainment. To simulate real-world retrieval scenarios, we treat article headlines as queries and the corresponding article bodies as passages. As the dataset lacks explicit relevance judgments, we adopt a heuristic supervision approach: each headline is assumed to be relevant to its associated article. To validate this assumption, we manually examined a random subset of query-passage pairs and confirmed high topical alignment between headlines and their articles. We also removed duplicates using \texttt{MD5} hashing and reformatted the data into an MS MARCO-style passage retrieval format. This results in approximately \texttt{45K} query-passage pairs. We split the dataset into training and test sets, reserving \texttt{10\%} for evaluation. The split is stratified by category to ensure balanced representation across all six news domains.

\subsection{Implementation Details}
\label{details}

\heading{Amharic embedding models.} We trained our Amharic embedding models on a single \texttt{A100 40GB} GPU for \texttt{4} epochs using the Sentence Transformer Trainer from the \texttt{sentence-transformers} Python library.\footnote{\url{https://pypi.org/project/sentence-transformers/}} Training was performed with a learning rate of \texttt{5e-5}, batch size \texttt{128}, cosine learning rate scheduler, and the multiple negatives ranking loss (MNRL) for optimization.

\heading{Sparse retrieval baselines.} For BM25-based retrieval, we used the BM25Retriever from the \texttt{LlamaIndex} framework.\footnote{\url{https://docs.llamaindex.ai/en/stable/examples/retrievers/bm25_retriever/}}


\heading{Dense retrieval baseline.}
We implemented ColBERT using the \texttt{PyLate} library~\cite{PyLate},\footnote{\url{https://github.com/lightonai/pylate}} adapting it for Amharic using the \textit{RoBERTa-Medium-Amharic} encoder model. The model was trained with a learning rate of \texttt{1e-5} and batch size \texttt{32}, using eight negative samples drawn from the top \texttt{150} passages retrieved by our \textit{RoBERTa-Medium-Amharic-Embed model}.

\heading{Fine-tuning multilingual models.}
We fine-tuned the Snowflake-Arctic-Embed model on Amharic query–passage pairs for \texttt{4} epochs using the AdamW optimizer with a learning rate of \texttt{2e-5}, batch size \texttt{128}, and a linear warmup ratio of \texttt{0.1}. We applied a weight decay of \texttt{0.01} and used a cosine scheduler with warmup.

\heading{Evaluation metrics.}
We evaluate retrieval effectiveness using standard ranking metrics in \ac{IR}:
\begin{enumerate*}[label=(\roman*)]
    \item {MRR@$k$} (mean reciprocal rank): evaluates the average inverse rank of the first relevant passage.

    \item {NDCG@$k$} (normalized discounted cumulative gain): assesses ranking quality with graded relevance and logarithmic position discounting; in our case, it is computed using binary relevance labels.

    \item {Recall@$k$}: measures how often relevant passages appear within the top-\(k\) retrieved results.

\end{enumerate*}

%% file: sections/Results.tex
\section{Experimental Evaluation and Results}
\label{sec:res}
In this section we present our empirical evaluation, which is structured around the following research questions:

\begin{table*}[t]
    \centering
    \setlength{\tabcolsep}{3pt} 
    \begin{tabular}{cl@{}rccccc}
        \toprule
        &&&&& \multicolumn{3}{c}{Recall} \\
        \cmidrule{6-8}
        &Model & Params & MRR@10 & NDCG@10 & @10 & @50 & @100 \\
        \midrule
        \multirow{4}{*}{\rotatebox[origin=c]{90}{\parbox[c]{1.8cm}{\centering\em Multilingual\\models}}}&
        gte-modernbert-base & 149M & 0.019 & 0.023 & 0.033 & 0.051 & 0.067 \\
        & gte-multilingual-base & 305M & 0.600 & 0.638 & 0.760 & 0.851 & 0.882 \\
        & multilingual-e5-large-instruct & 560M & 0.672 & 0.709 & 0.825 & 0.911 & 0.931 \\
        & snowflake-arctic-embed-l-v2.0 & 568M & 0.659 & 0.701 & 0.831 & 0.922 & 0.942 \\
        \midrule
        \multirow{3}{*}{\rotatebox[origin=c]{90}{\em Ours}} 
        & BERT-Medium-Amharic-Embed & 40M & 0.682 & 0.720 & 0.843 & 0.931 & 0.954 \\
        & RoBERTa-Medium-Amharic-Embed & 42M & 0.735 & 0.771 & 0.884 & 0.955 & 0.971 \\
        & RoBERTa-Base-Amharic-Embed & 110M & \textbf{0.775}\rlap{$^\dagger$} & \textbf{0.808}\rlap{$^\dagger$} & \textbf{0.913}\rlap{$^\dagger$} & \textbf{0.964}\rlap{$^\dagger$} & \textbf{0.979}\rlap{$^\dagger$} \\
        \bottomrule
    \end{tabular}
\caption{
Performance comparison on the Amharic Passage Retrieval Dataset between our Amharic-optimized embedding models and state-of-the-art multilingual dense retrieval baselines, all based on a bi-encoder architecture. 
The multilingual models \textit{snowflake-arctic-embed-l-v2.0} and \textit{multilingual-e5-large-instruct} originate from Arctic Embed 2.0~\cite{yu2024arctic} and Multilingual E5 Text Embeddings~\cite{wang2024multilingual}, respectively. 
Best results are shown in \textbf{bold}. 
Statistically significant improvements ($p < 0.05$) over the strongest multilingual baseline are marked with $^\dagger$, based on a paired t-test.
}
\label{tab:multilingual_vs_am}
\end{table*}

\begin{enumerate}[label=\textbf{RQ\arabic*}, nosep, leftmargin=*]
    \item How well do Amharic-optimized embeddings improve ranking accuracy compared to general-purpose multilingual embedding models? (Section~\ref{sec:multilingual_vs_amharic})
    
    \item How do different retrieval paradigms compare in effectiveness, establishing a benchmark for Amharic passage retrieval? (Section~\ref{sec:sparse_vs_dense})
    
    \item How does tokenization quality, particularly subword segmentation, impact retrieval effectiveness in morphologically rich, low-resource languages like Amharic? (Section~\ref{sec:Fertility})
    
    \item To what extent does the base model size influence retrieval performance in late interaction models for low-resource settings like Amharic? (Section~\ref{sec:ablation})
\end{enumerate}

\subsection{Evaluating Amharic Embeddings Against Multilingual Baselines}
\label{sec:multilingual_vs_amharic}

We investigate whether Amharic-optimized embedding models offer tangible advantages over general-purpose multilingual models in ranking Amharic passages. Table~\ref{tab:multilingual_vs_am} compares three Amharic-specific models with four multilingual baselines using standard IR metrics. Across the board, Amharic-optimized models outperform multilingual counterparts, often with fewer parameters. The best-performing multilingual model, \textit{Snowflake-Arctic-Embed} (\texttt{568M} parameters), achieves 0.659 MRR@10, whereas \textit{RoBERTa-Base-Amharic-Embed} (\texttt{110M} parameters) reaches 0.775, reflecting a 17.6\% relative gain. Similar improvements are observed in NDCG@10 (0.808 vs.\ 0.701) and Recall@10 (0.913 vs.\ 0.831), demonstrating consistent gains across top- and mid-rank positions. Notably, \textit{RoBERTa-Medium-Amharic-Embed} (\texttt{42M}) outperforms all multilingual models in MRR@10 and Recall@10 despite being over \texttt{13}$\times$ smaller than \textit{Snowflake-Arctic-Embed}. This finding underscores that scaling multilingual models does not necessarily translate into better retrieval performance for low-resource languages.

These findings emphasize three key insights:
\begin{enumerate*}[label=(\roman*)]
    \item {Tokenization alignment matters:} Amharic-optimized models better preserve word boundaries, reducing subword fragmentation and improving semantic matching (see Section~\ref{sec:Fertility}).
    \item {Parameter efficiency matters:} Amharic-specific models achieve superior performance with significantly fewer parameters.
    \item {Language-specific adaptation outperforms brute-force scaling:} Fine-tuning on monolingual data yields greater benefit than applying large multilingual encoders out-of-the-box.
\end{enumerate*}

\subsection{Benchmarking Sparse vs.\ Dense Retrieval for Amharic IR}
\label{sec:sparse_vs_dense}

We compare sparse and dense retrieval paradigms to establish strong baselines for Amharic passage retrieval. As shown in Table~\ref{tab:sparse_vs_dense_baselines}:
\begin{enumerate*}[label=(\roman*)]
    \item BM25 serves as a competitive sparse baseline, achieving 0.657 MRR@10 and 0.774 Recall@10, reaffirming its relevance in low-resource settings.

\item Dense retrieval models outperform this baseline across all evaluation metrics. The bi-encoder model \textit{RoBERTa-Base-Amharic-Embed} improves upon BM25 with 0.775 MRR@10 and 0.913 Recall@10, highlighting the benefits of Amharic-specific embeddings. Its Recall@100 score of 0.979 also indicates strong coverage across larger candidate sets.

\item The best-performing system is \textit{ColBERT-RoBERTa-Base-Amharic}, a late interaction model built on the same Amharic encoder. By incorporating token-level interactions, it significantly enhances precision, achieving 0.843 MRR@10 and 0.939 Recall@10, a 28.31\% relative improvement in MRR over BM25. It also surpasses the bi-encoder at top and mid ranks (e.g., Recall@50: 0.972 vs.\ 0.964), while maintaining parity at Recall@100 (0.979). These results highlight the complementary strengths of Amharic-specific encoders and interaction-aware architectures.
\end{enumerate*}
Overall, these findings demonstrate the effectiveness of dense retrieval methods, particularly late interaction models like ColBERT, when paired with language-specific pretraining. Both dense systems benefit from Amharic-optimized encoders, underscoring the importance of tailoring retrieval architectures to the linguistic characteristics of morphologically rich, low-resource languages.

\begin{table*}[t]
    \centering
    \setlength{\tabcolsep}{4pt} 
    \begin{tabular}{llccccc}
        \toprule
        &&&& \multicolumn{3}{c}{Recall} \\
        \cmidrule{5-7}
        Type & Model & MRR@10 & NDCG@10 & @10 & @50 & @100 \\
        \midrule
        Sparse retrieval & BM25-AM & 0.657 & 0.682 & 0.774 & 0.847 & 0.871 \\
        Dense retrieval & RoBERTa-Base-Amharic-Embed & 0.775 & 0.808 & 0.913 & 0.964 & 0.979 \\
        Dense retrieval & ColBERT-RoBERTa-Base-Amharic & \textbf{0.843}\rlap{$^\dagger$} & \textbf{0.866}\rlap{$^\dagger$} & \textbf{0.939}\rlap{$^\dagger$} & \textbf{0.972}\rlap{$^\dagger$} & {0.979} \\
        \bottomrule
    \end{tabular}
    \caption{
    Performance of retrieval models on the Amharic Passage Retrieval Dataset. 
    \textit{ColBERT-RoBERTa-Base-Amharic} is a late interaction model that builds on the \textit{RoBERTa-Base-Amharic-Embed} encoder. 
    Best results are shown in \textbf{bold}. Statistically significant improvements ($p < 0.05$) over the strongest baseline are marked with $^\dagger$, based on a paired t-test.
    }
    \label{tab:sparse_vs_dense_baselines}
\end{table*}

\subsection{Tokenization Quality and Retrieval Performance}
\label{sec:Fertility}

\begin{figure}[!t]  
    \centering
    \includegraphics[width=\linewidth]{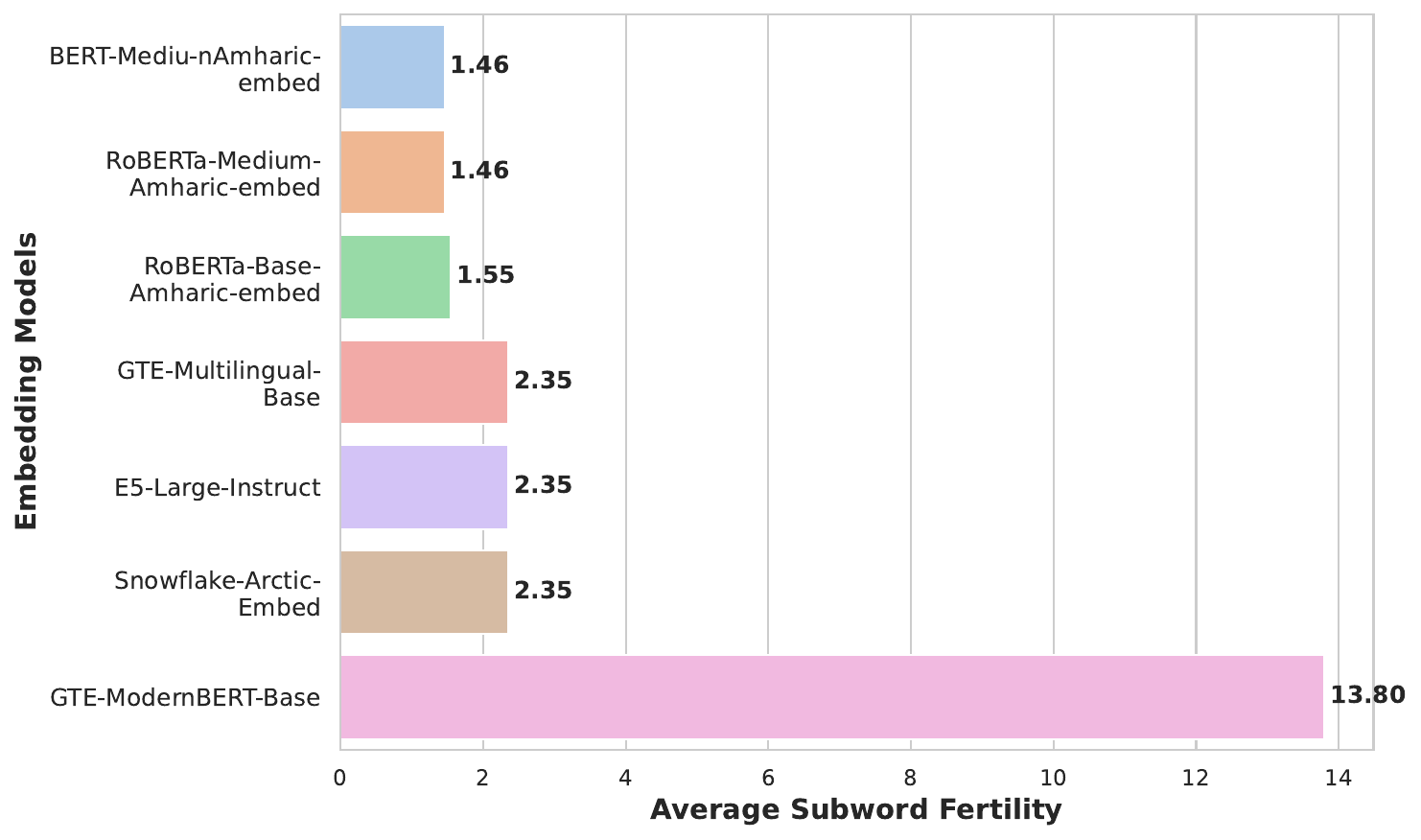} 
    \caption{Average subword fertility across embedding models. Lower fertility indicates better alignment with word boundaries, while higher fertility suggests excessive segmentation, which can harm retrieval accuracy.}
    \label{fig:Fertility}
\end{figure}

This section investigates how tokenization quality, particularly subword segmentation, impacts retrieval effectiveness in morphologically rich, low-resource languages, using Amharic as a case study. We focus on subword fertility, defined as the average number of subword tokens per word~\cite{TokenizationQuality}, as a key indicator of tokenization quality. Figure~\ref{fig:Fertility} presents fertility scores across various embedding models, based on a representative subset of \texttt{10k} Amharic passages.

Excessive subword segmentation (i.e., high fertility) increases computational overhead and fragments semantic representations, which degrades retrieval accuracy~\cite{ali-etal-2024-tokenizer}. For example:
\begin{enumerate*}[label=(\roman*)]
\item \textit{gte-modernbert-base} exhibits the highest fertility (13.80) and the weakest retrieval performance (MRR@10 = 0.019), demonstrating the detrimental effects of poor tokenization. In contrast, Amharic-optimized models such as \textit{RoBERTa-Base-Amharic-Embed} achieve the lowest fertility (1.46) and the highest MRR@10 (0.775), indicating better alignment between tokenization and linguistic structure.
\item Among multilingual models, \textit{snowflake-arctic-embed-l-v2.0} demonstrates moderate fertility (2.35) and the best performance in its category (MRR@10 = 0.659), likely benefiting from its large parameter size (\texttt{568M}). However, it still underperforms relative to much smaller Amharic-specific models, suggesting that model size alone cannot compensate for tokenization inefficiencies.
\end{enumerate*}

\begin{figure}[t]
\centering
\includegraphics[clip,trim=1mm 2mm 2mm 1mm, width=\linewidth]{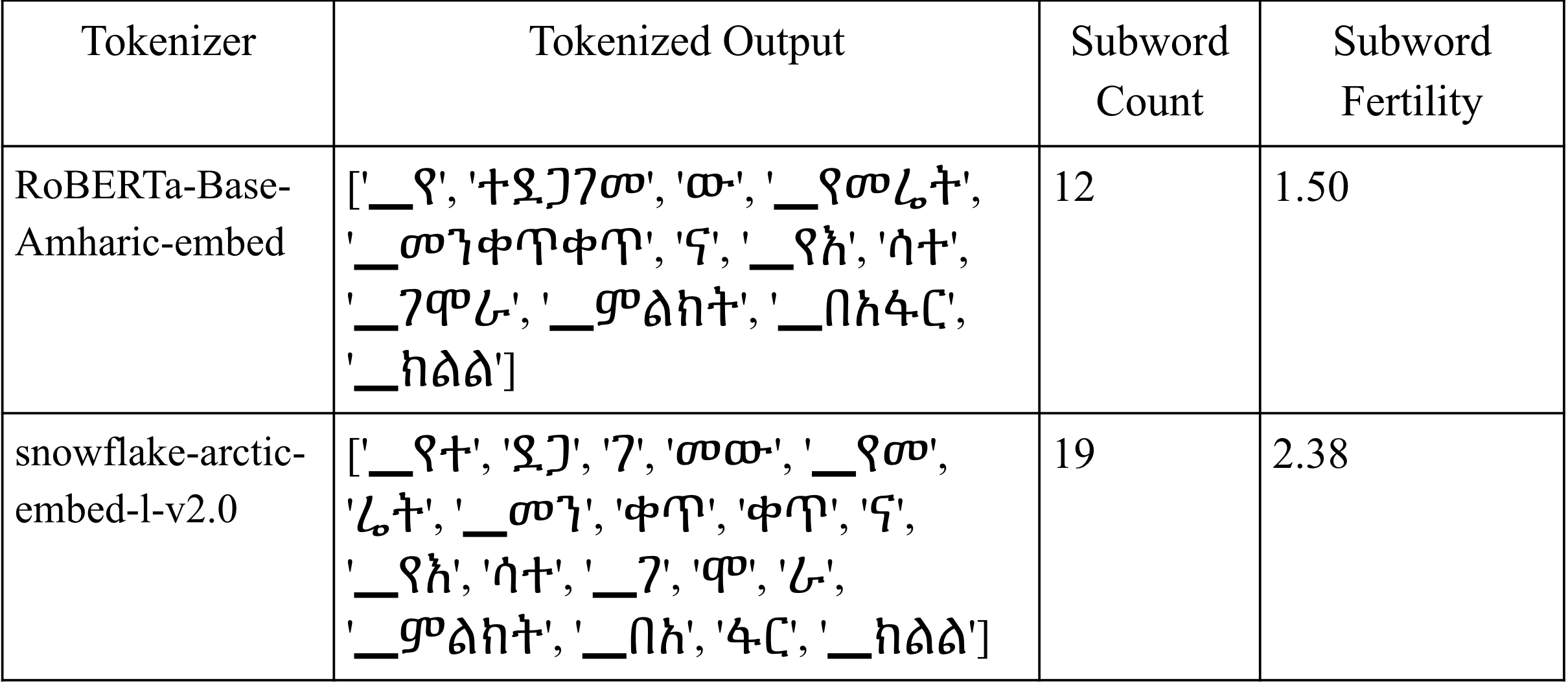}
\caption{
Subword tokenization comparison for a representative Amharic sentence. \textit{RoBERTa-Base-Amharic-Embed} produces more compact and linguistically meaningful tokens than \textit{snowflake-arctic-embed-l-v2.0}, reducing subword fragmentation and improving semantic representation quality.
}
\label{fig:amharic-table}
\end{figure}

These findings are consistent with prior work~\cite{Turkish, ali-etal-2024-tokenizer}, reinforcing the critical role of tokenizer alignment, particularly in morphologically complex languages, in improving computational efficiency and downstream retrieval performance.
%
%
To further illustrate this issue, Figure~\ref{fig:amharic-table} presents a qualitative comparison of subword tokenization for a representative Amharic sentence. We contrast the segmentation behavior of the best-performing Amharic-specific model (\textit{RoBERTa-Base-Amharic-Embed}) with that of the strongest multilingual model (\textit{snowflake-arctic-embed-l-v2.0}). The Amharic-specific model generates fewer and more linguistically coherent tokens, which likely contributes to its superior retrieval performance.

\subsection{Model Size vs.\ Performance in Late Interaction Retrieval}
\label{sec:ablation}

\begin{figure}[!t]  
    \centering
    \includegraphics[width=1.0\linewidth]
    {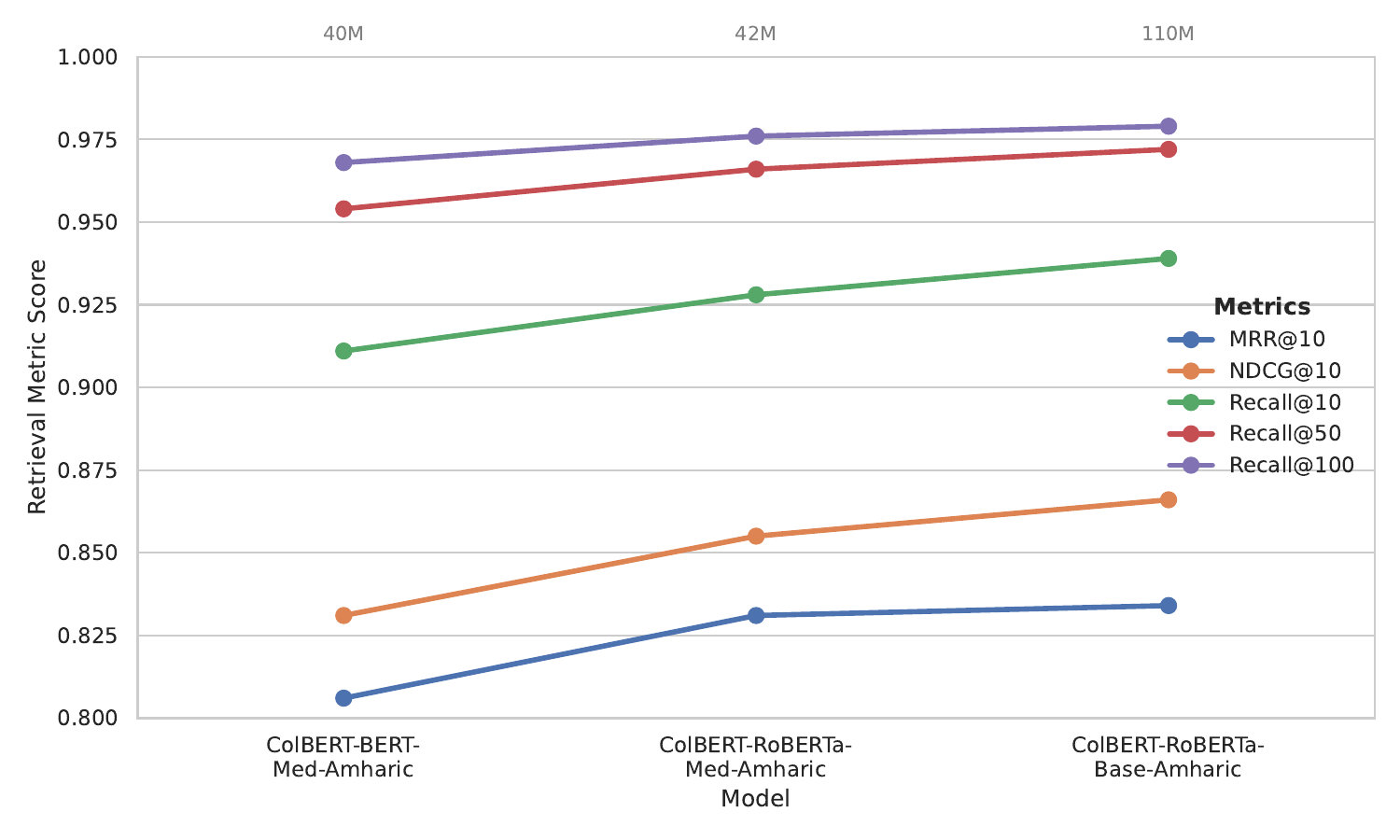}
\caption{
Effect of base model size on ColBERT performance in Amharic passage retrieval. 
The figure presents retrieval effectiveness across five ranking metrics for ColBERT models initialized with different Amharic base encoders. 
Lines connect performance metrics per model to highlight comparative trends. 
}

    \label{fig:ablation_colbert}
\end{figure}

We investigate the trade-off between model size and retrieval effectiveness by comparing three Amharic encoder models within a late interaction framework using ColBERT: \textit{BERT-Medium-Amharic}, \textit{RoBERTa-Medium-Amharic}, and \textit{RoBERTa-Base-Amharic}. Figure~\ref{fig:ablation_colbert} summarizes performance across five retrieval metrics, highlighting how encoder size influences ranking accuracy and recall in Amharic passage retrieval.

\begin{enumerate*}[label=(\roman*)]
\item \textit{ColBERT-RoBERTa-Base-Amharic} (\texttt{110M}) achieves the best overall performance (MRR@10: 0.843, NDCG@10: 0.866, Recall@10: 0.939), suggesting that scaling up the encoder benefits token-level retrieval, likely due to increased representational capacity.

\item \textit{RoBERTa-Medium-Amharic} (\texttt{42M}) remains highly competitive (MRR@10: 0.831, Recall@10: 0.928), achieving a 1.5\% relative performance difference from its larger counterpart while being 62\% smaller, demonstrating strong efficiency in resource-constrained scenarios.

\item \textit{BERT-Medium-Amharic} (\texttt{40M}) also performs strongly (MRR@10: 0.806), showing that compact models remain viable for retrieval in low-resource settings.
\end{enumerate*}

While larger models boost ColBERT’s performance, well-optimized medium-sized encoders strike a more favorable balance between accuracy and efficiency, making them ideal for compute-constrained, low-resource settings.


%

\subsection{Fine-Tuning Multilingual Models with Amharic Supervision}

While our primary comparison focuses on zero-shot multilingual models, we also investigate the impact of retrieval-specific supervised fine-tuning. To this end, we fine-tune the strongest multilingual baseline, \textit{Snowflake-Arctic-Embed} (\texttt{568M} parameters), using Amharic query–passage pairs. The resulting model, \textit{snowflake-arctic-embed-l-v2.0-AM}, shows substantial performance improvements: MRR@10 increases from 0.659 to 0.827, and Recall@10 rises from 0.831 to 0.942 (Table~\ref{tab:finetuned_multilingual}).

These results highlight two key insights:
\begin{enumerate*}[label=(\roman*)]
\item {Even large, multilingual embedding models are suboptimal} for low-resource retrieval tasks when used \mbox{out-of-the-box}.
\item Retrieval-specific supervision with in-language data significantly improves ranking effectiveness, especially at top ranks (MRR@10: +25.5\%).
\end{enumerate*}
This underscores the importance of task-aligned and language-specific adaptation. Notably, retrieval fine-tuning enhances semantic alignment more effectively than general-purpose multilingual pretraining, even without modifying the underlying architecture.

\begin{table*}[t]
\centering
\setlength{\tabcolsep}{6pt}
\begin{tabular}{lccccc}
\toprule
\textbf{Model} & \textbf{MRR@10} & \textbf{NDCG@10} & \multicolumn{3}{c}{\textbf{Recall}} \\
\cmidrule(lr){4-6}
& & & @10 & @50 & @100 \\

\midrule
snowflake-arctic-embed-l-v2.0 & 0.659 & 0.701 & 0.831 & 0.922 & 0.942 \\

{snowflake-arctic-embed-l-v2.0-AM} & \textbf{0.827}$^\dagger$ & \textbf{0.855}$^\dagger$ & \textbf{0.942}$^\dagger$ & \textbf{0.977}$^\dagger$ & \textbf{0.985}$^\dagger$ \\
\bottomrule
\end{tabular}
\caption{
Effect of Amharic-specific fine-tuning on multilingual retrieval performance. 
\textit{snowflake-arctic-embed-l-v2.0-AM} denotes the fine-tuned variant trained on Amharic passage-level supervision.
$^\dagger$ indicates statistically significant improvements ($p < 0.05$) over the zero-shot version, based on a paired t-test.
}
\label{tab:finetuned_multilingual}
\end{table*}

\subsection{Key Challenges in Amharic Passage Retrieval}
While Table~\ref{tab:multilingual_vs_am} shows that Amharic-optimized models like \textit{RoBERTa-Base-Amharic-Embed} consistently outperform multilingual baselines, several persistent challenges reveal the underlying complexity of Amharic IR:
\begin{enumerate*}[label=(\roman*)]
\item {Morphological complexity:} Amharic’s templatic morphology results in diverse word forms. Despite improved tokenization in language-specific models, subword over-segmentation, especially for inflected or compound words, still fragments semantics and limits retrieval accuracy.

\item {Data scarcity:} Amharic models are pretrained on just \texttt{300M} tokens, far fewer than for high-resource languages. This restricts generalization, particularly for rare terms or specialized domains, and contributes to residual retrieval errors even in strong models.

\item {Evaluation noise:} The Amharic passage retrieval dataset lacks human-annotated relevance labels, relying instead on headline–article pairs as heuristic signals. While practical, this weak supervision introduces noise and limits the granularity of relevance modeling.

\item Qualitative observations: Manual inspection of top-ranked outputs shows that Amharic-optimized dense models generally retrieve more contextually appropriate content. However, even the best models struggle with negation, temporal shifts, and nuanced entailment. For instance, given the query \emph{“Was the planned protest not held?”}, the model retrieved a passage stating \emph{“The planned protest was held,”} ranking it highly despite the semantic contradiction. Sparse models, by contrast, often favor surface-level keyword overlap (e.g., matching on \emph{“protest”}), yet fail to account for polarity or temporal context. These observations highlight that retrieval effectiveness still hinges on capturing deeper semantic and discourse-level nuances, an open challenge in low-resource settings.

\end{enumerate*}

These challenges are further discussed in the limitations (Section~\ref{sec:limit}) and illustrated with qualitative error analysis in (Appendix~\ref{sec:error}), highlighting fundamental issues in low-resource IR and emphasizing the need for better tokenization, richer training corpora, and curated evaluation benchmarks.

%% file: sections/Conclusion.tex
\section{Conclusion}
\label{sec:conclusion}

We introduced dense retrieval models and established the first systematic benchmark for Amharic passage retrieval. Our models consistently outperform multilingual baselines, underscoring the importance of linguistic adaptation for morphologically rich, low-resource languages. We also show that tokenization quality, especially subword fertility, significantly impacts retrieval performance: compact segmentations improve ranking accuracy, while over-segmentation harms semantic alignment. Our main experiments use the Amharic Passage Retrieval Dataset (derived from AMNEWS using heuristic labels), and we include supplementary results on 2AIRTC in the appendix. 

However, both datasets present evaluation challenges: the former lacks gold-standard relevance judgments, and the latter has incomplete labeling. These limitations underscore the need for more robust evaluation resources and motivate future research directions.

To address these gaps, future work should focus on:  
\begin{enumerate*}[label=(\roman*)]
\item designing morphology-aware or byte-level tokenizers tailored to Amharic's templatic structure,
\item improving training with hard negative mining and curriculum-based strategies, and 
\item extending evaluation to document-level and multi-hop retrieval.
\end{enumerate*}
Creating a high-quality, human-annotated benchmark with expert-labeled relevance, dialect variation, and morphological features, through collaboration with local institutions will be critical for aligning IR systems with real-world Amharic information needs.



%% file: sections/acknowledgements.tex
\section*{Acknowledgements}
We are grateful to our reviewers for their thoughtful comments and insightful feedback, which helped us improve the quality of this work.

This work was partially supported by the Dutch Research Council (NWO) under project EINF-9550, with computations performed on the Snellius supercomputer (SURF). Additional support was provided by the Dutch Research Council (NWO) under project numbers 024.004.022, NWA.1389.20.183, and KICH3.LTP.20.006; and the European Union's Horizon Europe program under grant agreement No 101070212.

The content of this paper reflects the views of the authors and does not necessarily represent the official position of their affiliated institutions or sponsors.

%% file: sections/Limitations.tex
\section{Limitations}
\label{sec:limit}
\heading{Dataset and evaluation.} 
We rely on the Amharic Passage Retrieval Dataset (derived from AMNEWS), which lacks human-annotated relevance judgments. Our assumption that headlines reflect document relevance introduces weak supervision noise. Furthermore, the dataset’s limited scale constrains generalizability. Future work should consider collecting crowd-sourced labels or leveraging Amharic language models for automatic annotation to enhance evaluation fidelity.

\heading{Pretraining data.}  
Our Amharic base models were pre-trained on a relatively modest corpus of 300 million tokens from web, news, and social media sources. This is substantially smaller than the corpora used for high-resource language, e.g., English BERT (3.3B) and RoBERTa (30B). Such data limitations may affect model generalization and downstream retrieval performance.

\heading{Domain generalization.} 
The main experiments were conducted within the news domain. The effectiveness of our retrieval models in other domains (e.g., medical, legal, or technical) remains untested and would likely require further domain adaptation.

\heading{Tokenization and morphology.}  
Amharic’s templatic morphology poses tokenization challenges, which we analyze using subword fertility. However, our models do not incorporate explicit morphological analyzers, lemmatizers, or segmentation tools. Instead, we rely on standard tokenization and language-specific fine-tuning. Tokenization inconsistencies introduce over-segmentation, degrading semantic coherence and retrieval accuracy. These limitations open avenues for future work, including the integration of morphology-aware tokenizers, hybrid word–subword representations, and explicit linguistic preprocessing pipelines.

\heading{Fine-tuning strategy.}
We employed full-parameter fine-tuning to maximize retrieval effectiveness in our monolingual Amharic setup, where preserving multilingual capabilities was not a priority. While this approach yields strong performance, future work should explore parameter-efficient alternatives such as LoRA or lightweight adapters, especially in cross-lingual settings where model compactness and multilingual retention are essential.


%% file: sections/Ethical_considerations.tex
\section{Ethical Considerations}
Our study aims to improve passage retrieval for Amharic, a low-resource language. While our models show substantial performance gains, we acknowledge potential ethical concerns regarding data biases, fairness, and deployment risks.

\heading{Use of publicly available data.}
We use two public datasets: AMNEWS~\cite{am_news_data}, comprising news articles, and 2AIRTC~\cite{2AIRTC}, a TREC-style IR dataset. All data is publicly available, and no new data was collected, ensuring compliance with ethical standards.

\heading{Base models and pretraining data.}  
Our Amharic embeddings are derived from models pre-trained on \texttt{300M} tokens of publicly accessible Amharic web, news, and tweet data. We use existing checkpoints from Hugging Face and rely on their accompanying documentation for data provenance.

\heading{Bias and fairness considerations.}  
Like many datasets sourced from online news content, the AMNEWS dataset may contain inherent biases related to reporting styles, topic framing, and regional representation. Retrieval models trained on this dataset may inherit and reflect these biases, particularly for politically or socially sensitive topics. While our study does not explicitly mitigate bias, we recognize this as an important challenge and encourage future work on fairness-aware retrieval and debiasing strategies.

\heading{Algorithmic challenges in low-resource languages.}  
Amharic is a low-resource, morphologically rich language, making it susceptible to algorithmic disparities due to data sparsity and tokenization challenges. While we highlight these issues, our approach does not introduce direct mitigation techniques beyond language-specific fine-tuning. Future work should explore improved tokenization and linguistic adaptation methods to enhance retrieval fairness.

\heading{Responsible deployment and transparency.}  
We follow ACL’s ethical standards and stress that models should not be deployed in high-stakes applications without rigorous auditing. We support transparency in sharing model limitations and advocate for careful, informed use of our publicly released models and datasets.

We encourage the community to use our models and datasets responsibly, and to continue advancing equitable IR systems that serve linguistically diverse users.

%% file: sections/Appendix.tex
\appendix
\section*{Appendix}
\section{2AIRTC: Amharic Ad Hoc Information Retrieval Test Collection}
\label{sec:appendix}

2AIRTC~\cite{2AIRTC} is the first TREC-style test collection for Amharic Information Retrieval (IR), comprising \texttt{12,583} documents and \texttt{240} manually assessed search topics. Each topic includes a title, description, and narrative (in both Amharic and English), with relevance judgments provided in standard QREL format. The dataset spans diverse domains (e.g., news, religion, culture, politics) and includes full-length documents sourced from news outlets, Wikipedia, social media, and blogs.

\heading{Limitations of 2AIRTC.}  
Despite its foundational role, 2AIRTC presents several limitations that restrict its utility for robust and reproducible evaluation:

\begin{enumerate}[leftmargin=*,label=(\roman*)]
    \item \textbf{Incomplete relevance judgments:}  
    Many semantically relevant documents remain unjudged, particularly those retrieved by neural models relying on semantic similarity. This leads to underestimated performance, especially for recall-based metrics, and compromises evaluation reliability.

    \item \textbf{Lack of standardized baselines:}  
    The absence of published baselines or leaderboard comparisons limits reproducibility and makes it difficult to benchmark retrieval systems fairly across studies.
\end{enumerate}

\noindent These limitations underscore the need for updated, high-coverage Amharic IR benchmarks with exhaustive annotations and unified evaluation protocols to ensure fair, consistent, and progress-driving comparisons in future research.

\subsection{Generalization to 2AIRTC: Amharic-Specific vs.\ Multilingual Models}
\label{sec:generalization}

To assess the generalization capacity of retrieval models trained on the Amharic Passage Retrieval Dataset, we evaluate their zero-shot performance on 2AIRTC, the only publicly available TREC-style benchmark for Amharic ad hoc retrieval. Despite known limitations such as annotation sparsity, 2AIRTC provides a valuable secondary testbed to evaluate retrieval robustness beyond the news domain. Table~\ref{tab:zero} compares multilingual and Amharic-specific dense retrievers on this corpus.

Amharic-specific models, despite having significantly fewer parameters, demonstrate competitive generalization. For instance, \textit{RoBERTa-Base-Amharic-embed} achieves 0.770 NDCG@100 and 0.910 Recall@200, just one point below the strongest multilingual baseline (\textit{multilingual-e5-large-instruct}) while being over 5$\times$ smaller. This highlights the strength of compact, linguistically aligned models for retrieval in low-resource settings.

Interestingly, performance does not scale monotonically with model size. \textit{gte-multilingual-base} (305M) outperforms the larger \textit{snowflake-arctic-embed-l-v2.0} (568M), indicating that architecture and pretraining objectives can outweigh parameter count.

\vspace{1mm}
\noindent\textbf{Key Findings:}
\begin{enumerate}[leftmargin=*,label=(\roman*)]
\item {Language-specific models generalize effectively:} Despite smaller model size, Amharic-optimized models closely match multilingual systems, offering efficient and scalable alternatives for retrieval in low-resource languages.

    \item {Cross-benchmark variance reveals sensitivity to evaluation design:} Amharic-specific models outperform on the Amharic Passage Retrieval Dataset but achieve comparable rather than dominant performance on 2AIRTC. This reflects differences in domain and the impact of sparse or incomplete relevance annotations.
    
    \item {Dense models are disadvantaged by annotation sparsity:} Dense retrievers rely on semantic similarity, often surfacing relevant but unjudged content. The incomplete supervision in 2AIRTC penalizes these models on recall-based metrics, underestimating their true effectiveness.
\end{enumerate}

\noindent%
These results emphasize the utility of Amharic-specific models for retrieval in low-resource contexts, while also underscoring the need for more complete and semantically annotated benchmarks to fairly assess dense retrievers' performance across domains.

\begin{table*}[t]
    \centering
    \setlength{\tabcolsep}{2pt}  
    \begin{tabular}{l ccccc}
        \toprule
        &&&& \multicolumn{2}{c}{Recall}
        \\
        \cmidrule{5-6}
        Model & Params & 
        MRR@100 & NDCG@100 & @100 & @200 \\
        \midrule
        \multicolumn{6}{l}{\emph{Multilingual Models}} \\
        \midrule
        gte-modernbert-base & 149M & 0.046 & 0.017 & 0.021 & 0.033 \\
        gte-multilingual-base & 305M & 0.879 & 0.749 & 0.790 & 0.865 \\
        multilingual-e5-large-instruct & 560M & \textbf{0.905} & \textbf{0.808} & \textbf{0.853} & \textbf{0.911} \\
        snowflake-arctic-embed-l-v2.0 & 568M & 0.876 & 0.781 & 0.830 & 0.897 \\
        \midrule
        \multicolumn{6}{l}{\emph{Ours}} \\
        \midrule
        BERT-Medium-Amharic-embed & \phantom{0}40M &  0.805 & 0.667 & 0.727 & 0.828 \\
        RoBERTa-Medium-Amharic-embed & \phantom{0}42M & 0.853 & 0.735 & 0.798 & 0.878 \\
        RoBERTa-Base-Amharic-embed & 110M & 0.861\rlap{$^\uparrow$}  & 0.770 & 0.830 & 0.910\rlap{$^\uparrow$} \\
     
        \bottomrule
    \end{tabular}
\caption{Performance comparison of Amharic-optimized and multilingual dense retrieval models, all based on a bi-encoder architecture, evaluated on the 2AIRTC dataset. The models snowflake-arctic-embed-l-v2.0 and multilingual-e5-large-instruct (Hugging Face model names) originate from Arctic Embed 2.0 \cite{yu2024arctic} and Multilingual E5 Text Embeddings \cite{wang2024multilingual}, respectively. The best-performing results are highlighted in \textbf{bold} and the second best in up-arrow \rlap{$^\uparrow$}.}      
\label{tab:zero}
\end{table*}

\subsection{Impact of Fine-Tuning on Cross-Domain Generalization}
\label{sec:finetune_2airtc}

To examine whether supervised fine-tuning improves cross-domain generalization, we evaluate \textit{snowflake-arctic-embed-l-v2.0-AM}, a multilingual model fine-tuned on the Amharic Passage Retrieval Dataset, on the 2AIRTC benchmark without any additional adaptation.

Table~\ref{tab:2airc_finetuned_multilingual} presents the results. The fine-tuned model improves recall at both @100 and @200, achieving the highest Recall@200 (0.923) with a +2.6 point gain. It also shows a statistically significant increase in NDCG@100 (0.795 vs.\ 0.781), though MRR@100 slightly decreases. These findings suggest that retrieval-specific supervision on Amharic queries may enhance semantic alignment even across structurally different corpora. However, given 2AIRTC’s known limitations, such as sparse relevance annotations these results should be interpreted as indicative rather than conclusive.

\begin{table*}[t]
\centering
\setlength{\tabcolsep}{6pt}
\begin{tabular}{lcccc}
\toprule
\textbf{Model} & \textbf{MRR@100} & \textbf{NDCG@100} & \textbf{Recall@100} & \textbf{Recall@200} \\
\midrule
snowflake-arctic-embed-l-v2.0 & \textbf{0.876} & 0.781 & 0.830 & 0.897 \\
\textit{snowflake-arctic-embed-l-v2.0-AM} & 0.865 & \textbf{0.795}$^\dagger$ & \textbf{0.856}$^\dagger$ & \textbf{0.923}$^\dagger$ \\
\bottomrule
\end{tabular}
\caption{
Effect of Amharic domain-specific fine-tuning on cross-domain retrieval performance. 
\textit{snowflake-arctic-embed-l-v2.0-AM} is fine-tuned on AMNEWS and evaluated on 2AIRTC. 
$^\dagger$ indicates statistically significant improvements ($p < 0.05$) over the zero-shot baseline.}
\label{tab:2airc_finetuned_multilingual}
\end{table*}

 \subsection{ColBERT with Amharic-Specific Backbones on 2AIRTC}
\label{sec:colbert-2airtc}

We report the retrieval performance of three ColBERT variants equipped with Amharic-specific encoder backbones on the 2AIRTC dataset. All models were trained on the Amharic Passage Retrieval Dataset and evaluated zero-shot on 2AIRTC. Table~\ref{tab:colbert_retrieval_results} summarizes results across standard ranking metrics. Due to known limitations in 2AIRTC, including incomplete relevance judgments and annotation sparsity, we refrain from drawing strong conclusions and present these results as indicative for completeness.

\begin{table*}[t]
    \centering
    \setlength{\tabcolsep}{5pt}
    \begin{tabular}{l cc cc cc}
        \toprule
        \textbf{Model} 
        & \multicolumn{2}{c|}{\textbf{MRR}} 
        & \multicolumn{2}{c|}{\textbf{NDCG}} 
        & \multicolumn{2}{c}{\textbf{Recall}} \\
        \cmidrule(lr){2-3} \cmidrule(lr){4-5} \cmidrule(lr){6-7}
        & @100 & @200 
        & @100 & @200 
        & @100 & @200 \\
        \midrule
        ColBERT-BERT-Medium-Amharic & 0.907 & 0.907 & 0.823 & \textbf{0.842}\rlap{$^\dagger$} & 0.880 & \textbf{0.930}\rlap{$^\dagger$} \\
        ColBERT-RoBERTa-Medium-Amharic & 0.909 & 0.909 & 0.831 & 0.840 & 0.886 & 0.917 \\
        ColBERT-RoBERTa-Base-Amharic & \textbf{0.919}\rlap{$^\dagger$} & \textbf{0.919}& \textbf{0.834}\rlap{$^\dagger$} & 0.838 & \textbf{0.887}\rlap{$^\dagger$} & 0.906 \\
        \bottomrule
    \end{tabular}
    \caption{
        Retrieval performance of ColBERT models trained with different Amharic encoder backbones, evaluated at @100 and @200 cutoffs for MRR, NDCG, and Recall on 2AIRC dataset. 
        \textit{ColBERT-BERT-Medium-Amharic-AM} uses a medium-sized BERT encoder trained on the Amharic passage retrieval dataset; 
        \textit{ColBERT-RoBERTa-Medium-Amharic} uses a medium RoBERTa encoder trained on the same corpus; 
        \textit{ColBERT-RoBERTa-Base-Amharic} uses a larger RoBERTa base encoder finetuned for Amharic. 
        Best results are marked in \textbf{bold} and statistically significant differences ($p<0.05$) are indicated with $^\dagger$.
    }
    \label{tab:colbert_retrieval_results}
\end{table*}

\subsection{Toward Robust Benchmarks for Amharic Information Retrieval}
\label{sec:future_directions}

Although this study provides strong baselines for Amharic dense retrieval, the limitations of 2AIRTC, particularly its small query pool (240 topics) and sparse, sometimes inconsistent relevance annotations, significantly hinder its utility for rigorous evaluation. These limitations especially penalize dense models, which often retrieve semantically relevant but unjudged documents, leading to underreported performance on recall-oriented metrics.
To advance Amharic IR evaluation and support more reliable model development, we recommend the following future directions:
\begin{itemize}[leftmargin=*,nosep]
    \item \textbf{Refine and expand 2AIRTC:} Improve annotation quality and coverage through iterative assessments, leveraging expert review, crowdsourcing, or semi-automated labeling to address incompleteness and inconsistency.
    \item \textbf{Develop morphology-aware retrieval methods:} Introduce tokenization and matching techniques suited to Amharic’s templatic morphology, such as lemmatization or hybrid subword–word representations.
    \item \textbf{Enhance query modeling:} Apply Amharic-specific language models for query expansion and pseudo-relevance feedback to mitigate vocabulary mismatch and improve semantic coverage.
    \item \textbf{Establish multi-dataset evaluation standards:} Benchmark systems across across diverse Amharic retrieval datasets to assess robustness and generalizability, enabling more comprehensive evaluations and reproducible progress.
\end{itemize}
We hope future efforts will establish larger, expert-annotated testbeds that capture Amharic’s linguistic diversity, enabling more faithful and equitable IR system development.

\section{Amharic Passage Retrieval Dataset Limitations and Qualitative Error Analysis}
\subsection{Dataset Limitations}

While Section~\ref{sec:appendix} discusses 2AIRTC, here we focus on the Amharic Passage Retrieval Dataset used in our main experiments, constructed by pairing news headlines with their corresponding articles.
Each headline is treated as a query and its article as a relevant passage. While these headlines often serve as effective proxies for user queries, they are inherently editorial and concise, crafted to capture attention rather than to reflect authentic information-seeking behavior. This introduces a distributional gap between training-time queries and real-world user intent, which may limit generalization to practical retrieval scenarios.
Moreover, the dataset lacks explicit relevance judgments or user interaction signals (e.g., clicks, ratings). Negative examples are generated by sampling non-matching articles, but these may still be topically related or semantically similar. This can introduce label noise, weakening the learning signal during contrastive training. To address these gaps, future work should:
\begin{itemize}[leftmargin=*,nosep]
    \item Incorporate curated or user-derived queries (e.g., search logs or community Q\&A),
    \item Employ better hard negative mining strategies, and
    \item Collect human-annotated relevance labels for robust evaluation.
\end{itemize}

\subsection{Qualitative Error Analysis}
\label{sec:error}
To complement our quantitative evaluation, we conducted a small-scale qualitative analysis to better understand retrieval behaviors. We manually inspected top-ranked passages for selected queries across both sparse and dense systems. Amharic-optimized dense models generally retrieved semantically relevant content, often capturing broader meanings beyond exact keyword matches. In contrast, sparse models like BM25 tended to prioritize surface-level term overlap, sometimes surfacing passages that were topically misaligned despite lexical similarity.

One notable failure pattern involved the handling of negation. Dense models, despite their semantic capabilities, frequently retrieved similar or identical passages for both affirmative and negated versions of a query, failing to reflect the semantic reversal. This indicates that current Amharic embeddings may inadequately model negation, likely due to limited exposure to such constructs during pretraining.

\begin{figure*}[t]
\centering
\includegraphics[width=\linewidth]{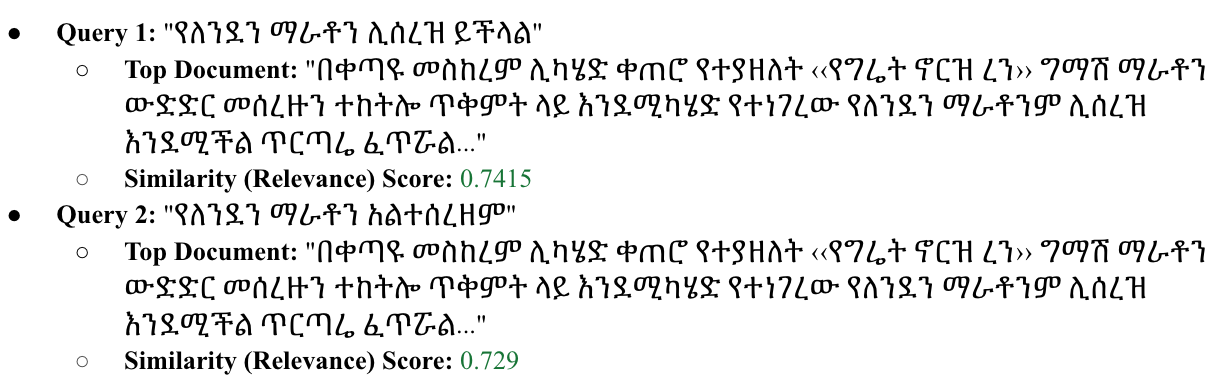}
\caption{Negation failure case: The model retrieves the same top passage for both a positive (Query 1) and a negated (Query 2) version of the query, with comparable similarity scores. This reflects a lack of semantic sensitivity to negation.}
\label{fig:negation}
\end{figure*}

Figure~\ref{fig:negation} illustrates this issue: despite the presence of negation in Query 2, the model ranks the same passage as for the affirmative Query 1, with nearly identical similarity scores. This suggests insufficient sensitivity to fine-grained semantic shifts like polarity reversal.
A broader set of such examples is provided in our \href{https://github.com/kidist-amde/amharic-ir-benchmarks/blob/main/notebooks/error_analysis_embedding_models.ipynb}{Python notebook}, available in the public GitHub repository.

\section{Hyperparameter Sensitivity}
\label{sec:hyperparam_sensitivity}

\begin{figure}[t]
\centering
\includegraphics[width=\linewidth]{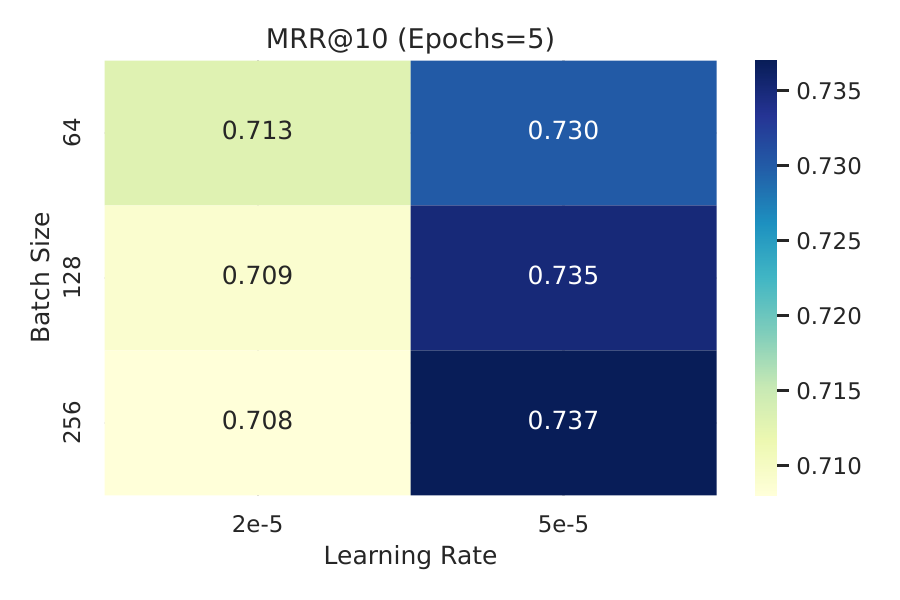}
\caption{MRR@10 scores with 5 training epochs. The best performance (\texttt{0.737}) is achieved with learning rate \texttt{5e-5} and batch size \texttt{256}. Higher learning rates consistently improve ranking quality across all batch sizes.}

\label{fig:heatmap_mrr10}
\end{figure}

\begin{figure}[t]
\centering
\includegraphics[width=\linewidth]{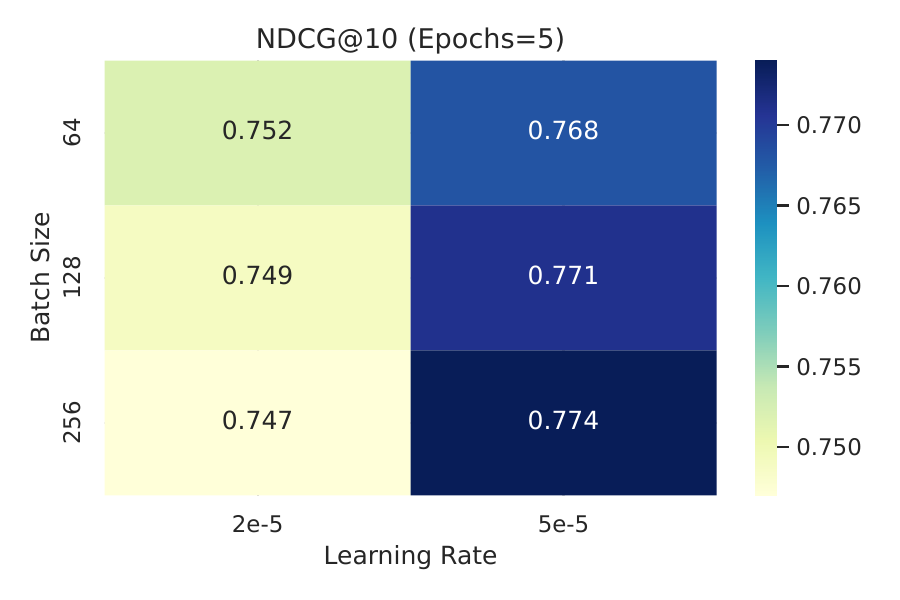}
\caption{NDCG@10 scores with 5 training epochs. Peak score (\texttt{0.774}) occurs at \texttt{5e-5} learning rate and batch size \texttt{256}. Larger batch sizes generally benefit from more aggressive learning.}
\label{fig:heatmap_ndcg10}
\end{figure}

\begin{figure}[t]
\centering
\includegraphics[width=\linewidth]{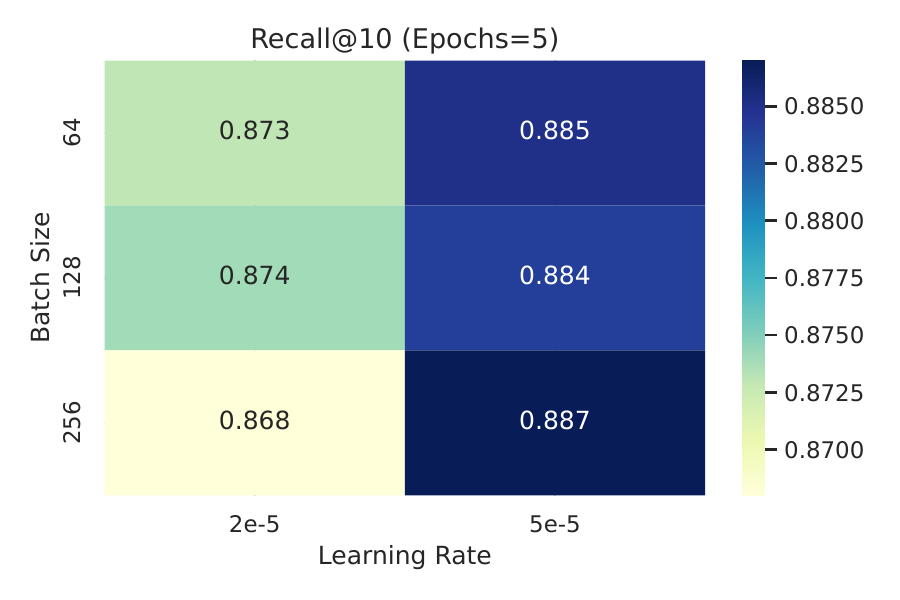}
\caption{Recall@10 under 5 training epochs. Maximum recall (\texttt{0.887}) is observed at \texttt{5e-5}/256. Performance improves steadily with training duration and a higher learning rate.}
\label{fig:heatmap_recal10}
\end{figure}

\begin{figure}[t]
\centering
\includegraphics[width=\linewidth]{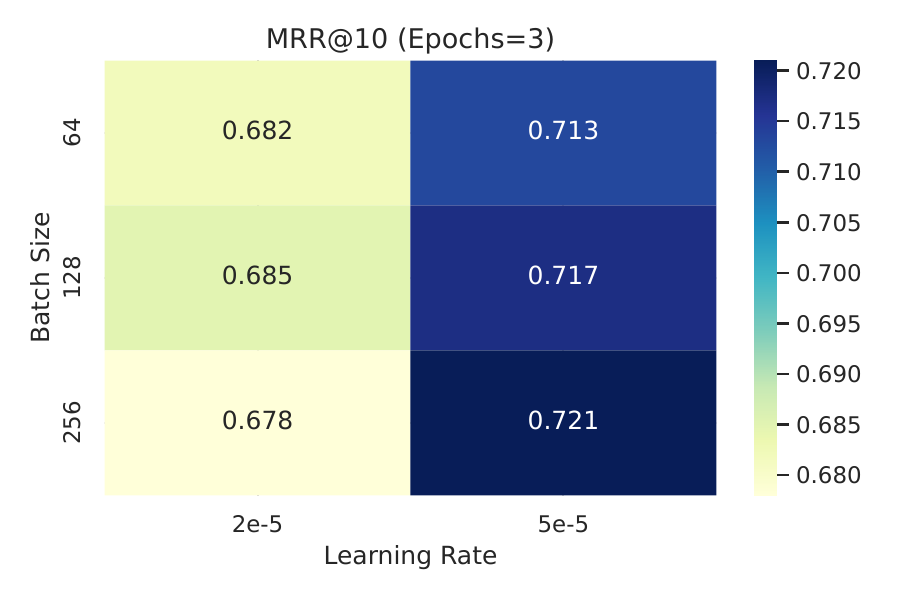}
\caption{MRR@10 with 3 training epochs. Best score (\texttt{0.721}) is attained at \texttt{5e-5}/256. Shorter training limits performance, but learning rate remains a strong influence.}
\label{fig:heatmap3_mrr10}
\end{figure}

\begin{figure}[t]
\centering

    \centering
    \includegraphics[width=\linewidth]{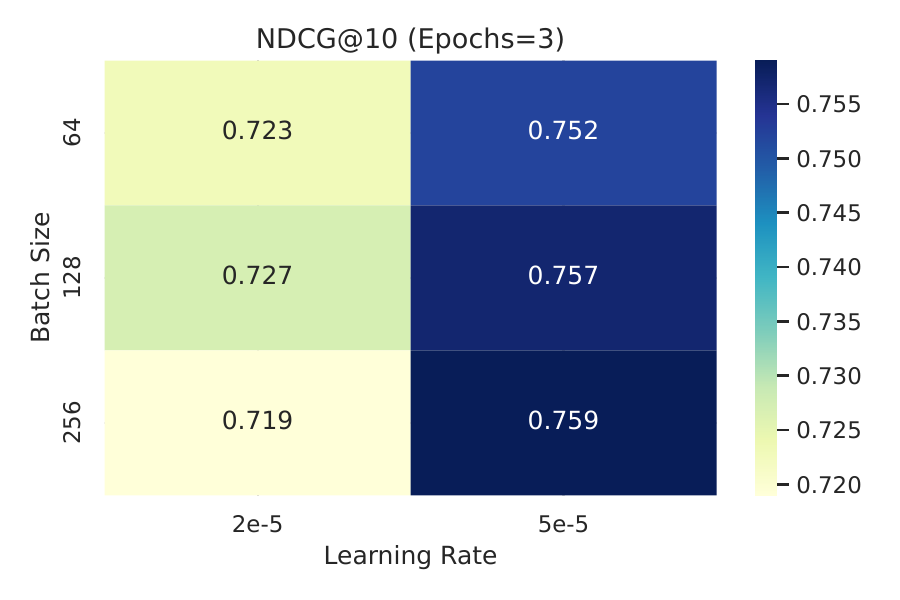}
    \caption{NDCG@10 with 3 training epochs. Performance is highest at \texttt{5e-5}/128, and all batch sizes benefit from higher learning rates.}
    \label{fig:heatmap3_ndcg10}
\end{figure}

\begin{figure}[t]

    \centering
    \includegraphics[width=\linewidth]{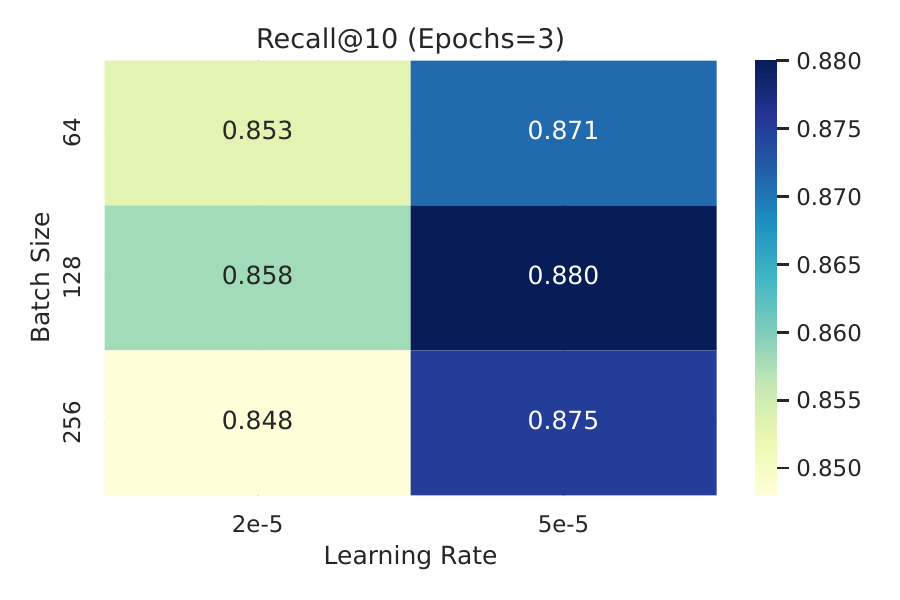}
    \caption{Recall@10 with 3 training epochs. The best score (\texttt{0.880}) is reached at \texttt{5e-5}/128, with higher learning rates consistently outperforming \texttt{2e-5}.}
    \label{fig:heatmap3_recal10}
\end{figure}

We conduct a grid search over learning rate, batch size, and training epochs using \textit{RoBERTa-Medium-Amharic-embed} to analyze the impact of hyperparameters on retrieval effectiveness. Figures~\ref{fig:heatmap_mrr10}--\ref{fig:heatmap3_recal10} present six heatmaps showing MRR@10, NDCG@10, and Recall@10 under two epoch settings (\textit{3} and \textit{5}). The results highlight that:
\begin{enumerate*}[label=(\roman*)]
\item increasing training epochs from 3 to 5 yields consistent improvements across all metrics. For example, with a learning rate of \texttt{5e-5} and batch size \texttt{256}, MRR@10 improves from \texttt{0.721} to \texttt{0.737}, and Recall@10 rises from \texttt{0.875} to \texttt{0.887}. 

\item Among learning rates, \texttt{5e-5} consistently outperforms \texttt{2e-5}, especially at larger batch sizes.

\item Batch size shows mild impact overall, with stable or slightly improved performance as size increases. The best overall configuration, \texttt{5e-5} learning rate, \texttt{256} batch size, and \texttt{5} epochs, achieves the top scores across all metrics, emphasizing the benefits of sustained training with a moderately aggressive learning rate.
\end{enumerate*}

These trends highlight that while batch size offers some flexibility, retrieval quality is more sensitive to learning rate and training duration.

\clearpage

\vspace*{10cm}

\vspace*{10cm}